\documentstyle[twocolumn,pra,aps]{revtex}
\begin{document}
\draft
\title{Efficient scheme for quantum entanglement, quantum information transfer, and
quantum gate with three-level SQUID qubits in cavity QED}
\author{Chui-Ping Yang$^{1,2}$, Shih-I Chu$^{2}$, and Siyuan Han$^{1}$}
\address{$^1$Department of Physics and Astronomy, University of Kansas,\\
Lawrence, Kansas 66045}
\address{$^2$Department of Chemistry, University of Kansas, and Kansas Center\\
for Advanced Scientific Computing, Lawrence, Kansas 66045}
\maketitle

\begin{abstract}
A novel scheme is proposed for realizing quantum entanglement, quantum
information transfer and a set of universal quantum gates with
superconducting-quantum-interference-device (SQUID) qubits in cavity QED. In
the scheme, the two logical states of a qubit are the two lowest levels of
the SQUID. An intermediate level of the SQUID is utilized to facilitate
coherent control and manipulation of quantum states of the qubits. The
method presented here does not create finite intermediate-level population
or cavity-photon population during the operations. Thus, decoherence due to
spontaneous decay from the intermediate levels is minimized and the
requirement on the quality factor of the cavity is greatly loosened.
\end{abstract}

\pacs{PACS number{s}: 03.67.Lx, 03.65.-w, 74.50.+r, 85.25.Dq, 42.50.Dv}


Cavity QED has been extensively studied to implement quantum information
processing (QIP) with a variety of physical systems such as atoms, ions,
quantum dots and Josephson junctions [1-6]. A well-known reason for this is
that compared with those non-cavity proposals where significant overhead is
needed for coupling distant qubits, the cavity-based schemes is preferable
since the cavity mode acts as a ``bus'' that can mediate long-range fast
interaction between any qubits, which enables one to perform two-qubit gates
involving any desired pair of qubits.

Recently, a scheme has been proposed for obtaining a complete set of
universal quantum gates, quantum information transfer, and entanglement with
superconductor quantum interference devices (SQUIDs) in cavity QED [7].
Technically speaking, the SQUID-cavity QED scheme may be among the most
promising candidates for demonstrating QIP because placing SQUIDs at desired
positions is straightforward in a cavity and superconducting qubits have
been demonstrated to have relatively long decoherence time [8-10]. In Ref.
[7], the gates were performed by inducing transitions to the intermediate
level $\left| a\right\rangle $ [see Fig. 1(a)] via microwave pulse and
cavity field. However, though the cavity mode is not populated during the
operation, the population of the SQUIDs in the intermediate levels is
non-zero. Thus, the operation must be done within a much shorter time than
the energy relaxation time of the intermediate level to maintain coherence.
Another key point is that the operation in [7] requires rapid adjustments of
level spacings of SQUIDs, which might be undesirable in experiment.

In this letter, we propose a significantly improved approach to achieve
entanglement, information transfer and universal gates with three-level $%
\Lambda $-type SQUID qubits in cavity QED. The new method has three major
advantages: (a) during the gate operations, the intermediate level is
unpopulated and thus decoherence induced by spontaneous emission from the
intermediate level, is greatly suppressed; (b) no transfer of quantum
information between the SQUIDs and the cavity is required, i.e., the cavity
field is only virtually excited and thus the requirement on the quality
factor of the cavity is relaxed; (c) there is no need to adjust the level
spacings during the operation.

Let us first introduce the Hamiltonian of a SQUID qubit coupled to a
single-mode cavity field and a classical microwave pulse with ${\bf B}_{\mu
w}({\bf r},t)={\bf B}_{\mu w}({\bf r})$ cos $\omega _{\mu w}t.$ Here, ${\bf B%
}_{\mu w}({\bf r})$ is the amplitude of the magnetic component and $\omega
_{\mu w}$ is the carrier frequency. The qubits considered in this letter are
rf SQUIDs each consisting of a Josephson tunnel junction in a
superconducting loop (typical size of an rf SQUID is on the order of 10 $\mu
m-$100 $\mu m$). The Hamiltonian of an rf SQUID (with junction capacitance $%
C $ and loop inductance $L$) can be written in the usual form 
\begin{equation}
H_s=\frac{Q^2}{2C}+\frac{\left( \Phi -\Phi _x\right) ^2}{2L}-E_J\cos \left(
2\pi \frac \Phi {\Phi _0}\right) ,
\end{equation}
where $\Phi $, the magnetic flux threading the ring, and $Q$, the total
charge on the capacitor, are the conjugate variables of the system (with the
commutation relation $\left[ \Phi ,Q\right] =i\hbar $), $\Phi _x$ is the
static (or quasistatic) external flux applied to the ring, and $E_J$ $\equiv
I_c\Phi _0/2\pi $ is the maximum Josephson coupling energy ($I_c$ is the
critical current of the junction and $\Phi _0=h/2e$ is the flux quantum).

The quantized Hamiltonian of the cavity mode is given by $H_c=\hbar \omega
_c\left( c^{+}c+1/2\right) ,$ where $c^{+}$ and $c$ are the photon creation
and annihilation operators; and $\omega _c$ is the frequency of the cavity
mode.

Consider a $\Lambda $-type configuration formed by the two lowest levels and
an excited level of the SQUID, denoted by $\left| 0\right\rangle ,\left|
1\right\rangle $ and $\left| a\right\rangle $ with energy eigenvalues $E_0,$ 
$E_1,$ and $E_a,$ respectively [Fig. 1(a)]. For the sake of concreteness, we
choose the following device and control parameters: $C=90$ fF$,L=100$ pH$%
,I_c=3.75$ $\mu $A$,\Phi _x=0.4995$ $\Phi _0$ for the SQUID qubit in the
rest of this letter. Suppose that the coupling of $\left| 0\right\rangle
,\left| 1\right\rangle $ and $\left| a\right\rangle $ with the other levels
via the cavity mode and the microwave is negligible (e.g., by adjusting
cavity size, microwave frequency, or level spacings of the SQUID). Under
this assumption, it is easy to find that when the cavity mode is coupled to
the $\left| 0\right\rangle \leftrightarrow \left| a\right\rangle $
transition but far-off resonant with the $\left| 0\right\rangle
\leftrightarrow \left| 1\right\rangle $ and $\left| 1\right\rangle
\leftrightarrow \left| a\right\rangle $ transitions, and when the microwave
pulse is coupled to the $\left| 1\right\rangle \leftrightarrow \left|
a\right\rangle $ transition while far-off resonant with the $\left|
0\right\rangle \leftrightarrow \left| 1\right\rangle $ and $\left|
0\right\rangle \leftrightarrow \left| a\right\rangle $ transitions, the
Hamiltonian of the system can be written as: 
\begin{eqnarray}
H &=&E_0\sigma _{00}+E_1\sigma _{11}+E_a\sigma _{aa}+\hbar \omega _cc^{+}c 
\nonumber \\
&&\ \ \ \ \ \ +\hbar (gc^{+}\sigma _{0a}+h.c.)+\hbar \left( \Omega
e^{i\omega _{\mu w}t}\sigma _{1a}+h.c.\right) ,
\end{eqnarray}
where $g$ is the coupling constant between the cavity mode and the $\left|
0\right\rangle \leftrightarrow \left| a\right\rangle $ transition; $\Omega $
is the Rabi-flopping frequency corresponding to the $\left| 1\right\rangle
\leftrightarrow \left| a\right\rangle $ transition; and $\sigma _{ij}=\left|
i\right\rangle \left\langle j\right| $ ($i,j=0,1,a$). The expressions of $g$
and $\Omega $ are given by [7] 
\begin{eqnarray*}
g &=&\frac 1L\sqrt{\frac{\omega _c}{2\mu _0\hbar }}\left\langle 0\right|
\Phi \left| a\right\rangle \int_S{\bf B}_c({\bf r})\cdot d{\bf S}, \\
\Omega &=&\frac 1{2L\hbar }\left\langle 1\right| \Phi \left| a\right\rangle
\int_S{\bf B}_{\mu w}({\bf r})\cdot d{\bf S,}
\end{eqnarray*}
where $S$ is any surface that is bounded by the SQUID ring, ${\bf r}$ is the
position vector on $S$, and ${\bf B}_c({\bf r})$ is the magnetic component
of the normal mode of the cavity.

Consider a situation in which the cavity mode is largely detuned from the $%
\left| 0\right\rangle \leftrightarrow \left| a\right\rangle $ transition ,
i.e., $\Delta _c=\omega _{a0}-\omega _c\gg g,$ and the microwave pulse is
largely detuned from the $\left| 1\right\rangle \leftrightarrow \left|
a\right\rangle $ transition, i.e., $\Delta _{\mu w}=\omega _{a1}-\omega
_{\mu w}\gg \Omega ,$ where $\omega _{a0}=(E_a-E_0)/\hbar $ and $\omega
_{a1}=(E_a-E_1)/\hbar $ [Fig. 1(a)]. Under this condition, the intermediate
level $\left| a\right\rangle $ can be adiabatically eliminated [11,12].
Thus, the effective Hamiltonian in the interaction picture becomes [11,12] 
\begin{eqnarray}
H_i &=&\hbar [-\frac{g^2}{\Delta _c}c^{+}c\sigma _{00}-\frac{\Omega ^2}{%
\Delta _{\mu w}}\sigma _{11}  \nonumber \\
&&\ \ -g_{eff}e^{i\delta t}c\sigma _{01}^{+}-g_{eff}e^{-i\delta
t}c^{+}\sigma _{01}],
\end{eqnarray}
where $\sigma _{01}=\left| 0\right\rangle \left\langle 1\right| ,$ $\sigma
_{01}^{+}=\left| 1\right\rangle \left\langle 0\right| ,$ $\delta =\Delta
_c-\Delta _{\mu w},$ and $g_{eff}=\frac{\Omega g}2(\frac 1{\Delta _c}+\frac 1%
{\Delta _{\mu w}}).$ The first two terms are ac-Stark shifts of the levels $%
\left| 0\right\rangle $ and $\left| 1\right\rangle $ induced by the cavity
mode and the microwave pulse, respectively. The last two terms are the
familiar Jaynes-Cummings interaction, describing the Raman coupling of the
two lowest levels of the SQUID.

{\it Effective Hamiltonian for two SQUID qubits in cavity. }To simplify
presentation, let us consider two identical SQUIDs $I$ and $II$ ( the method
is also applicable to non-identical SQUIDs). The two SQUIDs are coupled to
the same single-mode microwave cavity and each driven by a classical
microwave pulse ${\bf B}_{\mu w}^i({\bf r},t)={\bf B}_{\mu w}^i({\bf r})$
cos $\omega _{\mu w}t$ ($i=I,II$) [Fig. 1(c)]. The separation of the two
SQUIDs is assumed to be much larger than the linear dimension of each SQUID
ring in such a way that direct interaction between the two SQUIDs is
negligible. Also, suppose that the coupling of each SQUID to the cavity mode
is the same (this can be readily obtained by setting the two SQUIDs on two
locations ${\bf r}_1$ and ${\bf r}_2$ where the cavity-field magnetic
components ${\bf B}_c\left( {\bf r}_1,t\right) $ and ${\bf B}_c\left( {\bf r}%
_2,t\right) $ are the same). In this case, it is obvious that based on Eq.
(3), the Hamiltonian for the system in the interaction picture can be
written as 
\begin{eqnarray}
H_I &=&\sum_{i=I,II}\hbar [-\frac{g^2}{\Delta _c}c^{+}c\sigma _{00i}-\frac{%
\Omega ^2}{\Delta _{\mu w}}\sigma _{11i}]  \nonumber \\
&&\ \ \ \ -\hbar \sum_{i=I,II}[g_{eff}e^{i\delta t}c\sigma
_{01i}^{+}+g_{eff}e^{-i\delta t}c^{+}\sigma _{01i}].
\end{eqnarray}
Under the condition that $\delta \gg \frac{g^2}{\Delta _c},$ $\frac{\Omega ^2%
}{\Delta _{\mu w}},$ $g_{eff},$ there is no exchange of energy between the
SQUIDs and the cavity mode. The effective Hamiltonian is then given by
[13-16] 
\begin{eqnarray}
H_{eff} &=&\sum_{i=I,II}\hbar [-\frac{g^2}{\Delta _c}c^{+}c\sigma _{00i}-%
\frac{\Omega ^2}{\Delta _{\mu w}}\sigma _{11i}]  \nonumber \\
&&\ \ \ \ +\hbar \gamma [\sum\limits_{i=I,II}-c^{+}c\sigma
_{00i}+cc^{+}\sigma _{11i}  \nonumber \\
&&\ \ \ +\sigma _{01I}^{+}\sigma _{01II}+\sigma _{01I}\sigma _{01II}^{+}],
\end{eqnarray}
where the third and fourth terms describe the photon-number dependent Stark
shifts induced by the off-resonant Raman coupling, and the last two terms
describe the ``dipole'' coupling between the two SQUIDs mediated by the
cavity mode and the classical fields. The parameter $\gamma
=g_{eff}^2/\delta $ characterizes the strength of Stark shift and
inter-qubit coupling. If the cavity is initially in the vacuum state, then
the effective Hamiltonian reduces to 
\begin{eqnarray}
H_{eff} &=&-\sum_{i=I,II}\hbar \frac{\Omega ^2}{\Delta _{\mu w}}\sigma _{11i}
\nonumber \\
&&\ \ \ \ +\hbar \gamma [\sum\limits_{i=I,II}\sigma _{11i}+\sigma
_{01I}^{+}\sigma _{01II}+\sigma _{01I}\sigma _{01II}^{+}],
\end{eqnarray}
Note that the Hamiltonian (6) does not contain the operators of the cavity
mode. Thus, only the state of the SQUID system undergoes an evolution under
the Hamiltonian (6), i.e., no quantum information transfer occurs between
the SQUIDs and the cavity mode. Therefore, the cavity mode is virtually
excited.

The state $\left| 0\right\rangle _I\left| 0\right\rangle _{II}$ is
unaffected under the Hamiltonian (6). From (6), one can easily get the
following state evolution 
\begin{eqnarray}
\left| 0\right\rangle _I\left| 1\right\rangle _{II} &\rightarrow
&e^{-i\gamma ^{\prime }t}[\cos (\gamma t)\left| 0\right\rangle _I\left|
1\right\rangle _{II}-i\sin (\gamma t)\left| 1\right\rangle _I\left|
0\right\rangle _{II}],  \nonumber \\
\left| 1\right\rangle _I\left| 1\right\rangle _{II} &\rightarrow
&e^{-i2\gamma ^{^{\prime }}t}\left| 1\right\rangle _I\left| 1\right\rangle
_{II},
\end{eqnarray}
where $\gamma ^{\prime }=\gamma -\frac{\Omega ^2}{\Delta _{\mu w}}$. In the
following, we show that Eq. (7) can be used to create entanglement, to
implement quantum information transfer, and to perform quantum gates.

{\it Generation of entanglement.} The two logical states of each SQUID qubit
are represented by the two lowest energy states $\left| 0\right\rangle $ and 
$\left| 1\right\rangle .$ From (7), one can see that if the two SQUID qubits
are initially in the states $\left| 0\right\rangle _I$ and $\left|
1\right\rangle _{II}$, they will evolve to the following maximally entangled
state after an interaction time $\pi /(4\gamma )$ 
\begin{equation}
\left| \psi \right\rangle =\frac 1{\sqrt{2}}(\left| 0\right\rangle _I\left|
1\right\rangle _{II}-i\left| 1\right\rangle _I\left| 0\right\rangle _{II}),
\end{equation}
where the common phase factor $e^{-i\chi \pi /4}$ ($\chi =\gamma ^{\prime
}/\gamma $) has been omitted.

{\it Quantum information transfer}. Suppose that the SQUID qubit $I$ is the
original carrier of quantum information, which is in an arbitrary state $%
\alpha \left| 0\right\rangle +\beta \left| 1\right\rangle $. The quantum
state transfer from the qubit $I$ to the qubit $II$ initially in the state $%
\left| 0\right\rangle $ is described by 
\begin{equation}
(\alpha \left| 0\right\rangle _I+\beta \left| 1\right\rangle _I)\left|
0\right\rangle _{II}\rightarrow \left| 0\right\rangle _I(\alpha \left|
0\right\rangle _{II}+\beta \left| 1\right\rangle _{II}),
\end{equation}
which can be realized in the following two steps.

Step (i): Apply two microwave pulses to the two SQUIDs $I$ and $II$,
respectively, so that the states of the two SQUIDs undergo an evolution
under the Hamiltonian (6) for an interaction time $\pi /(2\gamma ).$

Step (ii): Perform a phase-shift $\left| 0\right\rangle \rightarrow
e^{-i(1+\chi )\pi /4}\left| 0\right\rangle $ while $\left| 1\right\rangle
\rightarrow e^{i(1+\chi )\pi /4}\left| 1\right\rangle $ on the SQUID qubit $%
II$ [17].

The states after each step of the above operations are listed below: 
\begin{eqnarray}
&&(\alpha \left| 0\right\rangle _I+\beta \left| 1\right\rangle _I)\left|
0\right\rangle _{II}\stackrel{\text{Step (i)}}{\longrightarrow }\left|
0\right\rangle _I[\alpha \left| 0\right\rangle _{II}+e^{-i(1+\chi )\pi
/2}\beta \left| 1\right\rangle _{II}]  \nonumber \\
&&\stackrel{\text{Step (ii)}}{\longrightarrow }e^{-i(1+\chi )\pi /4}\left|
0\right\rangle _I(\alpha \left| 0\right\rangle _{II}+\beta \left|
1\right\rangle _{II})
\end{eqnarray}
It is clear that the two-step operation transfers quantum information from
the SQUID qubit $I$ to the SQUID qubit $II$.

Single SQUID qubit operations can be achieved via various schemes [7,17,18].
In Ref. [17], it has been shown that by applying two microwave pulses to
induce two-photon Raman resonant transition between the qubit levels, any
single-SQUID-qubit logic operation can be realized, without real excitation
of the intermediate level. It is noted that during the present single-qubit
operation inside a cavity, the cavity mode can be decoupled from the qubits
without adjusting the qubits' level spacings. The reason for this is that
one can choose the frequencies of the applied microwave pulses so that
two-photon Raman resonant transition between the qubit levels $\left|
0\right\rangle $ and $\left| 1\right\rangle $ is satisfied, while the cavity
mode is highly detuned from either pulse [see Fig. 1 (b)].

{\it Quantum logical} {\it gates. }A non-trivial and universal two-qubit
controlled NOT (CNOT) can be realized by combining the Hamiltonian (6) with
single-qubit operations. We find that the CNOT gate $\left| i\right\rangle
_I\left| j\right\rangle _{II}\rightarrow \left| i\right\rangle _I\left|
i\oplus j\right\rangle _{II}$ ($i,j\in \{0,1\}$) acting on the two SQUID
qubits $I$ and $II$ can be achieved through the following unitary
transformations 
\begin{eqnarray}
{\cal U}_{CNOT} &=&{\cal H}_{II}^{-1}{\cal U}_I{\cal U}_{II}{\cal S}_I{\cal S%
}_{II}{\cal U}_{I,II}  \nonumber \\
&&\ \otimes {\cal \sigma }_{y_I}{\cal S}_I{\cal S}_{II}{\cal U}_{I,II}{\sc H}%
_{II}{\sc H}_I{\cal H}_{II},
\end{eqnarray}
where the common phase factor $e^{-i\chi \pi /4}$ is omitted, ${\cal U}%
_{I,II}$ is a two-SQUID-qubit joint unitary operation defined by ${\cal U}%
_{I,II}(\gamma t)=\exp [-\frac i\hbar H_{eff}t]$ with $\gamma t=\pi /4$, $%
{\cal \sigma }_y$ is the Pauli operator$,$ ${\cal S}$ results in a
single-qubit phase-shift $\left| 0\right\rangle \rightarrow e^{-i\chi \pi
/8} $ $\left| 0\right\rangle $ while $\left| 1\right\rangle \rightarrow
e^{i\chi \pi /8}\left| 1\right\rangle $, ${\cal U}_I=${\sc H}$_I^{-1}{\cal H}%
_I,$ ${\cal U}_{II}=${\sc H}$_{II}^{-1}{\cal H}_{II}^{-1},$ and ${\cal H},%
{\cal H}^{-1},${\sc H}$,${\sc H}$^{-1}$are the following Hadamard
transformations 
\begin{eqnarray}
{\cal H} &=&\frac 1{\sqrt{2}}\left( 
\begin{array}{cc}
1 & -1 \\ 
1 & 1
\end{array}
\right) ,{\sc H}=\frac 1{\sqrt{2}}\left( 
\begin{array}{cc}
1 & -i \\ 
-i & 1
\end{array}
\right) ,  \nonumber \\
{\cal HH}^{-1} &=&{\sc HH}^{-1}=\text{I}
\end{eqnarray}
in the single-qubit Hilbert subspace formed by $\left| 0\right\rangle
=(0,1)^{{\sc T}}$ and $\left| 1\right\rangle =(1,0)^{{\sc T}}.$

It is well known that at least three CNOT gates are needed [7] to construct
a two-qubit SWAP gate. Note that information transfer (9) is equivalent to a
transformation $\left| i\right\rangle _I\left| 0\right\rangle
_{II}\rightarrow \left| 0\right\rangle _I\left| i\right\rangle _{II}$ ($i\in
\left\{ 0,1\right\} $). Thus, to simplify the gate operation, a
two-SQUID-qubit SWAP $\left| i\right\rangle _I\left| j\right\rangle
_{II}\rightarrow \left| j\right\rangle _I\left| i\right\rangle _{II}$ ($%
i,j\in \left\{ 0,1\right\} $) can be realized through the following
procedure: 
\begin{eqnarray*}
\left| i\right\rangle _I\left| j\right\rangle _{II}\left| 0\right\rangle _a
&\rightarrow &\left| 0\right\rangle _I\left| j\right\rangle _{II}\left|
i\right\rangle _a\rightarrow \left| j\right\rangle _I\left| 0\right\rangle
_{II}\left| i\right\rangle _a \\
&\rightarrow &\left| j\right\rangle _I\left| i\right\rangle _{II}\left|
0\right\rangle _a.
\end{eqnarray*}

It is necessary to give a discussion on the effect of the Stark shift. From
(6), one can see that after omitting the Stark-shift terms, the Hamiltonian
(6) reduces to $H_{eff}^{\prime }=$ $\hbar \gamma [\sigma _{01I}^{+}\sigma
_{01II}+\sigma _{01I}\sigma _{01II}^{+}].$ For an arbitrary two-qubit state $%
\left| \psi \right\rangle =\alpha _0\left| 00\right\rangle +\alpha _1\left|
01\right\rangle +\alpha _2\left| 10\right\rangle +\alpha _3\left|
11\right\rangle $, the probability of the gate error caused by discarding
the Stark shift is given by 
\begin{eqnarray}
P_e &\equiv &1-{\cal F}  \nonumber \\
&=&4\sin ^2(\gamma ^{\prime }t/2)\left[ \left| \alpha _0\right| ^2(1-\left|
\alpha _0\right| ^2)+\left| \alpha _3\right| ^2(1-\left| \alpha _3\right|
^2)\right.  \nonumber \\
&&\ \ \left. +2\cos \gamma ^{\prime }t\left| \alpha _0\right| ^2\left|
\alpha _3\right| ^2\right]
\end{eqnarray}
where ${\cal F}=\left| \left\langle \psi \right| U^{-1}U^{\prime }\left|
\psi \right\rangle \right| ^2$ is the fidelity of the operation described by 
$U^{\prime }=\exp (-iH_{eff}^{\prime }t/\hbar ),$ in contrast to the
operation described by a real unitary operator $U=\exp (-iH_{eff}t/\hbar ).$
The equation (13) shows that for the special case of $\gamma ^{\prime
}t=2n\pi $ or $\alpha _0=$ $\alpha _3=0$, i.e., $\left| \psi \right\rangle
=\alpha _1\left| 01\right\rangle +\alpha _2\left| 10\right\rangle ,$ the
gate error is zero. However, it is noted that in general, one has $P_e\neq 0$%
, i.e., the gate operation will be affected by the Stark shift. In
particular, for $\alpha _0=$ $\alpha _3=\frac 1{\sqrt{2}},$ one has $%
P_e=\sin ^2(\gamma ^{\prime }t),$ which is unity for $\gamma ^{\prime
}t=(n+1/2)\pi $.

Finally, we show that parameters necessary for the experimental realization
of the proposed schemes are achievable. For the SQUIDs with the parameters
given above and with junction's damping resistance $R$ $>$ 1 G$\Omega $
[19], the level $\left| a\right\rangle $'s energy relaxation time $T_1\simeq 
\frac R{60\text{M}\Omega }\cdot \mu $s would be $\sim 15$ $\mu $s. The
transition frequency is $\omega _{a0}/(2\pi )\simeq $ 30 GHz. Hence, we
choose $\omega _c/(2\pi )=29.7$ GHz as the cavity-mode frequency. For a $%
10\times 1\times 1$ mm$^3$ cavity and a SQUID with a $50\times 50$ $\mu $m$%
^2 $ loop, a simple calculation shows that the coupling constant is $g\simeq 
$ $1.8\times 10^8$ s$^{-1}$, i.e., about $0.1\Delta _c.$ By choosing the
frequency and amplitude of the microwave pulse appropriately such that $%
\Delta _{\mu w}=10\Omega $ and $g=1.2\Omega $ for each SQUID, we have $%
\delta \simeq 10g_{eff}\simeq 3.0\times 10^8$ s$^{-1}$. Then the typical
time needed for the SQUID-cavity interaction is on the order of $T_{s-c}=\pi
\delta /(2g_{eff}^2)\simeq 0.5$ $\mu $s, which is much shorter than the
level $\left| a\right\rangle $'s effective decay time $T_1/P_a\geq 1.5\times
10^3$ $\mu $s for $T_1=15$ $\mu $s$,$ where $P_a\leq 0.01$ is the
occupational probability of the level $\left| a\right\rangle $ for the
present case of $\Delta _c=10g$ and $\Delta _{\mu w}=10\Omega $. The photon
lifetime is given by $T_c=Q_c/\omega _c$ where $Q_c$ is the quality factor
of the cavity. In the present case, the cavity has a probability $P_c\simeq
0.01$ of being excited during the operation. Thus, the effective decay time
of the cavity is $T_c/P_c\simeq 10$ $\mu $s $\gg T_{s-c}$ for $Q_c\simeq
2\times 10^4$, which is realizable as demonstrated by recent experiments
[20].

Note that the method described above does not require two SQUIDs with
identical parameters. In the case of non-identical SQUIDs $I$ and $II,$ one
has $\delta _I=\omega _{a0}^I-\omega _{a1}^I-\omega _c+\omega _{\mu w}^I$
and $\delta _{II}=\omega _{a0}^{II}-\omega _{a1}^{II}-\omega _c+\omega _{\mu
w}^{II}$, which can always be set to equal by adjusting the frequencies, $%
\omega _{\mu w}^I$ and $\omega _{\mu w}^{II},$ of the two microwave pulses
applied to the SQUIDs.

The present scheme has the following advantages: (i) During the operation,
the intermediate level is unpopulated and thus gate errors caused by energy
relaxation is greatly suppressed. (ii) The cavity field is virtually excited
and thus the required quality factor of the cavity is greatly loosened.
(iii). No tunneling between the qubit levels $\left| 0\right\rangle $ and $%
\left| 1\right\rangle $ is needed and thus the rate of spontaneous decay
from the level $\left| 1\right\rangle $ can be made negligibly small, by the
use of higher potential barrier between the two qubit levels. (iv) No
adjustment of level spacings is needed during logic operations, since the
qubit-qubit interaction required for the two-qubit operations is via the
cooperative actions of the cavity mode and the microwave pulses. (v) The
method can be extended to perform QIP on many SQUID qubits in a cavity,
because the cavity mode can mediate long-range coherent interaction between
SQUID qubits. Also, the proposal can be applied to any other type of solid
state qubits which have a $\Lambda $-type three-level configuration.

In summary, we have explicitly shown how quantum entanglement, quantum
information transfer, and universal quantum gates can be realized with SQUID
qubits in cavity. We stress that in our analysis, all Stark shift terms,
which may significantly affect the gate fidelity, are included. In addition,
we have shown that the method is feasible with experimentally demonstrated
qubit and cavity parameters. Thus, it provides a realistic approach for
robust quantum information processing with superconducting qubits, and we
hope that this work will stimulate further theoretical and experimental
activities in this emerging research field.

CPY is very grateful to Prof. Shi-Biao Zheng for many fruitful discussions
and very useful comments. This research was partially supported by National
Science Foundation (EIA-0082499), and AFOSR (F49620-01-1-0439), funded under
the Department of Defense University Research Initiative on Nanotechnology
(DURINT) Program and by the ARDA.

\begin{center}
{\large Figure Captions\\}
\end{center}

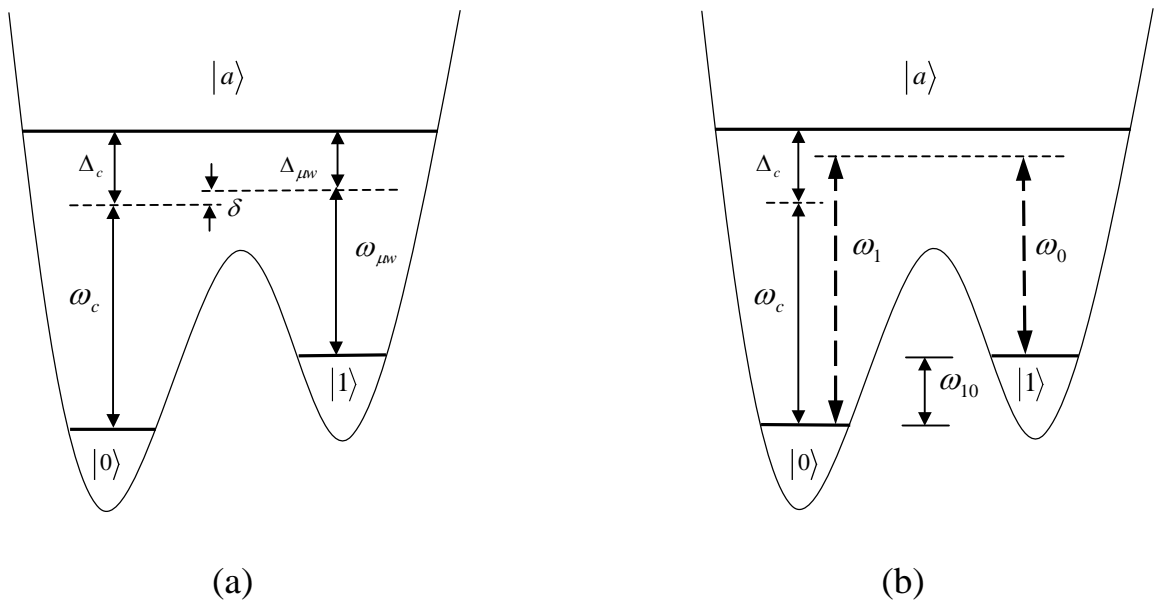
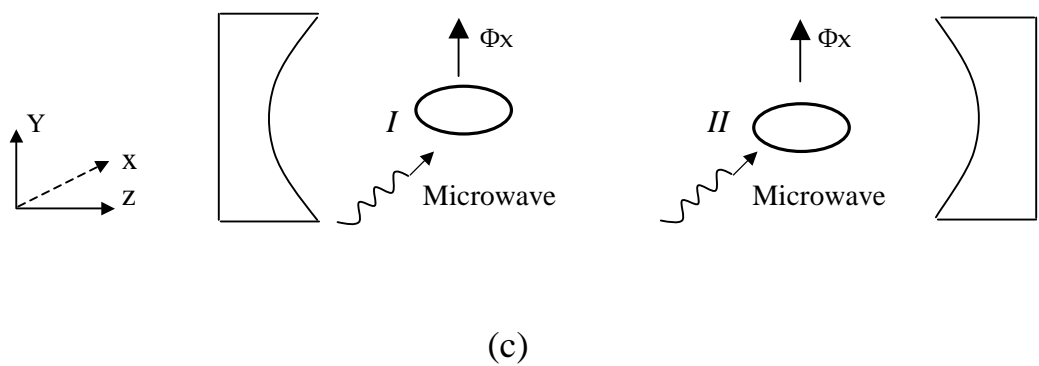
FIG. 1. (a) The potential and level diagram of an rf SQUID with a $\Lambda $%
-type three levels $\left| 0\right\rangle ,\left| 1\right\rangle $ and $%
\left| a\right\rangle $. The cavity field is detuned from the classical
microwave pulse by $\delta =\Delta _c-$ $\Delta _{\mu w}.$ (b) Illustration
of single-qubit operation. The two microwave pulses with frequencies $\omega
_0$ and $\omega _1$ are applied to induce two-photon Raman resonant
transition between the qubit levels $\left| 0\right\rangle $ and $\left|
1\right\rangle $ with $\omega _{10}=\omega _1$ $-$ $\omega _0,$ for the
purpose of single-qubit logic operation. (c) Schematic illustration of two
SQUIDs ($I,$ $II$) coupled to a single-mode cavity field and manipulated by
microwave pulses. The two SQUIDs are placed along the cavity axis (the $Z$
axis) and in the $X$-$Z$ plane. ${\bf B}_c,$ ${\bf B}_{\mu w\text{ }}^I$and $%
{\bf B}_{\mu w}^{II}$ are in $Y$ direction.

\end{document}